\documentclass[aps,prc,superscriptaddress,nofootinbib,showpacs,floatfix,singlecolumn]{revtex4}
\usepackage{url}
\usepackage{cancel}
\usepackage[colorlinks,linkcolor=blue,citecolor=blue,filecolor=black,urlcolor=blue]{hyperref}
\usepackage{epsfig,graphics}
\usepackage{graphicx}
\usepackage{dcolumn}
\usepackage{bm}
\usepackage[usenames]{color}
\usepackage{amssymb}
\usepackage{amsmath}
\usepackage{multirow}
\usepackage{float}
\usepackage{harpoon}
\usepackage{MnSymbol}
\usepackage{appendix}
\usepackage{color}
\usepackage{lineno}

\newcommand{\Rmnum}[1]{\expandafter\@slowromancap\romannumeral #1@}
\makeatother
\graphicspath{{./}{./fig}}


\newcommand{\sqrtsnn}{\mbox{$\sqrt{s_{\mathrm{NN}}}$}}

\newcommand{\Eref}{\mbox{$\eta_\mathrm{r}$}}

\newcommand{\lr}[1]{\left\langle #1\right\rangle}

\newcommand{\npart}{N_{\mathrm{part}}}
\newcommand{\nqp}{N_{\mathrm{qp}}}
\newcommand{\npartf}{N_{\mathrm{part}}^{\mathrm{F}}}
\newcommand{\npartb}{N_{\mathrm{part}}^{\mathrm{B}}}
\usepackage{CJK}
\hfuzz=\maxdimen
\begin{document}
\title{Longitudinal eccentricity decorrelations in heavy ion collisions}
\newcommand{\sbu}{Department of Chemistry, Stony Brook University, Stony Brook, NY 11794, USA}
\newcommand{\bnl}{Physics Department, Brookhaven National Laboratory, Upton, NY 11976, USA}
\newcommand{\sdua}{Institute of Frontier and Interdisciplinary Science, Shandong University, Qingdao, 266237, China}
\newcommand{\sdub}{Key Laboratory of Particle Physics and Particle Irradiation, Ministry of Education, Shandong University, Qingdao, Shandong, 266237, China}
 \author{Arabinda Behera}\affiliation{\sbu}\author{Maowu Nie}\email[]{maowu.nie@sdu.edu.cn}\affiliation{\sdua}\affiliation{\sdub}\author{Jiangyong Jia}\email[]{jiangyong.jia@stonybrook.edu}\affiliation{\sbu}\affiliation{\bnl}
\begin{abstract}
In heavy-ion collisions, the harmonic flow $V_n$ of final-state particles are driven by the eccentricity vector ${\mathcal{E}}_n$ that describe the shape of the initial fireball projected in the transverse plane. It is realized recently that the structure and shape of the fireball, and consequently the ${\mathcal{E}}_n$, fluctuate in pseudorapidity $\eta$ in a single event, ${\mathcal{E}}_n(\eta)$. This leads to eccentricity decorrelation between different $\eta$, driving the longitudinal flow decorrelations observed in the experiments. Using a Glauber model with a paramerterized longitudinal structure, we have estimated the eccentricity decorrelations and related them to the measured flow decorrelation coefficients for elliptic flow $n=2$ and triangular flow $n=3$. We investigated the dependence of eccentricity decorrelations on the choice of collision system in terms of the size, asymmetry and deformation of the nuclei. We found that these nuclear geometry effects lead to significant and characteristic patterns on the eccentricity decorrelations, which describe the measured ratios of the flow decorrelations between Xe+Xe and Pb+Pb collisions. These patterns can be searched for using existing experimental data at RHIC and the LHC, and if confirmed, they will provide a mean to improve our understanding of the initial state of the heavy-ion collisions. 
\end{abstract}
\maketitle
\section{Introduction}\label{sec:1}
Heavy ion collisions produce a quark-gluon plasma (QGP)~\cite{Shuryak:2014zxa,Busza:2018rrf} whose space-time evolution is well described by relativistic viscous hydrodynamics~\cite{Heinz:2013th,Jia:2014jca,Busza:2018rrf}. The QGP expansion converts the initial-state spatial anisotropies into final-state momentum anisotropies. These are characterized by Fourier expansion of azimuthal distribution of particle density, $dN/d\phi\propto 1+2\sum_{n=1}^\infty v_n \cos\,n(\phi-\Phi_n)$, where $v_n$ and $\Phi_n$ represent the amplitude and phase of the $n^{\mathrm{th}}$-order flow vector $V_n=v_n e^{{\textrm i}n\Phi_n}$. The $V_n$ reflects the hydrodynamic response of the produced medium to the $n^{\textrm{th}}$-order initial-state eccentricity vector~\cite{Gardim:2011xv,Niemi:2012aj}, denoted by ${\mathcal{E}}_n=\varepsilon_n {\mathrm e}^{{\textrm i}n\Phi^{\varepsilon}_n}$. Due to event-by-event (EbyE) density fluctuations in the initial state, the ${\mathcal{E}}_n$ and consequently the $V_n$ also fluctuate event to event. However, model calculations show that an approximate linear relation $V_n \propto {\mathcal{E}}_n$ is valid for $n=2$ (elliptic flow) and 3 (triangular flow) within a fixed centrality class, and the proportionality constant depends on the transport properties of the QGP~\cite{Luzum:2012wu,Teaney:2010vd,Gale:2012rq,Niemi:2012aj,Qiu:2012uy,Teaney:2013dta}.

Most previous efforts assumed that ${\mathcal{E}}_n$ and $V_n$ are boost invariant within a single event. But recent studies~\cite{Khachatryan:2015oea,Aaboud:2017tql} show significant fluctuations of harmonic flow along the longitudinal direction within the same event. This so called ``flow decorrelations'' appear as differences in flow magnitude ($v_n(\eta_1)\neq v_n(\eta_2)$) and its phase ($\Phi_n(\eta_1)\neq \Phi_n(\eta_2)$) along pseudorapidity ($\eta$). The origin can be attributed to the fact that the number of particle production sources and their transverse distribution fluctuates along $\eta$, which leads to longitudinal decorrelation of eccentricity vector in configuration space in a single event. For example, the number of forward-going and backward-going nucleon participants, $\npartf$ and $\npartb$, are not the same in a given event~\cite{Jia:2015jga,Jia:2014ysa}, and the corresponding eccentricity vectors ${\mathcal{E}}^{F}_n$ and ${\mathcal{E}}^{B}_n$ would also be different. Since these participants contribute differently to the final-state particles in the forward and backward rapidity, the eccentricity vector is closer to ${\mathcal{E}}^{F}_n$ (${\mathcal{E}}^{B}_n$) in the forward (backward) rapidity~\cite{Jia:2017kdq}. Hydrodynamic model simulations~\cite{Bozek:2010vz,Jia:2014ysa,Bozek:2015bna,Pang:2015zrq,Shen:2017bsr,Bozek:2017qir} show that the flow decorrelations reflect mainly the longitudinal structure of the initial state, and is insensitive to the viscosity of the QGP. Therefore, flow decorrelations serve as a unique probe for the early time dynamics of the heavy-ion collisions.

The first measurement of flow decorrelations was performed by the CMS Collaboration~\cite{Khachatryan:2015oea}, followed by a more detailed study by the ATLAS Collaboration~\cite{Aaboud:2017tql} in Pb+Pb collisions. Preliminary results have also been obtained at RHIC energies as well~\cite{Nie:2019bgd}. These results were described reasonably by several hydrodynamic model simulations with a 3D initial condition based on the lund-string picture~\cite{Bozek:2015bna,Pang:2015zrq}. Very recently, ATLAS also measured the flow decorrelations in the Xe+Xe system~\cite{Aad:2020gfz}. Compared with the Pb+Pb system, the decorrelation signal is observed to be larger for $v_2$, but smaller for $v_3$. Current hydrodynamic models~\cite{Pang:2018zzo,Wu:2018cpc} reproduce the $v_n$ in both systems but fail to describe simultaneously the centrality dependence of the $v_n$ decorrelations, which implies that the hydrodynamic models tuned to describe the transverse dynamics may not have the correct initial-state geometry in the longitudinal direction. However, in order to pin down exactly how to improve the description of 3D initial-state geometry, further systematic measurements and model studies in different collision systems are required.

In this paper, we explore longitudinal decorrelations of the initial-stage geometry using Glauber model simulations for different collision systems. We study the qualitative trends of the system size dependence in symmetric collision system, as well as the effects of the nuclear deformation and asymmetric collision system. We found that the decorrelations are sensitive to all these variations. 

\vspace*{-0.3cm}\section{Setup}\label{sec:2}
The longitudinal flow decorrelations are studied with a factorization ratio proposed by the CMS collaboration~\cite{Khachatryan:2015oea},
\begin{align}\label{eq:2}
r_{n}(\eta)=\frac{\langle V_n(-\eta)V_n^{*}(\Eref)\rangle}{\langle V_n(\eta)V_n^{*}(\Eref)\rangle} = \frac{\langle v_n(-\eta)v_n(\Eref)\,\cos\,n[\Phi_n(-\eta)-\Phi_n(\Eref)] \rangle}{\langle v_n(\eta)v_n(\Eref)\,\cos\,n[\Phi_n(\eta)-\Phi_n(\Eref)] \rangle}\approx 1- 2F_n \eta\;.
\end{align}
where $\Eref$ is a reference pseudorapidity range common to both the numerator and the denominator, and the average is done over events in a given centrality interval. Measurements show that $r_{n}(\eta)$ is an approximately linear function close to unity, and the slope parameter $F_n$ characterizes the strength of the decorrelation. 

Since flow vector and eccentricity vector are linearly correlated, $V_n\propto \mathcal{E}_n$, the $r_{n}(\eta)$ can be directly related to the initial eccentricity in spatial rapidity defined analogously to Eq.~\ref{eq:2}:
\begin{align}
\label{eq:e2}
r^{\rm s}_{n}(\eta) = \frac{\lr{\mathcal{E}_n (-\eta) \mathcal{E}_n^{*}(\Eref)}}{\lr{\mathcal{E}_n (\eta) \mathcal{E}_n^{*}(\Eref)}}
\end{align}
Hydrodynamic model calculations show that $r_{n}(\eta) \approx r^{\rm s}_{n}(\eta)$~\cite{Pang:2015zrq}, nearly independent of the value of shear viscosity in the final state.

Following our previous work~\cite{Jia:2017kdq}, the $\eta$ dependence of the eccentricity is estimated from the eccentricities of the forward-going and backward-going quark participants ${\mathcal{E}}_n^{\rm F}$ and ${\mathcal{E}}_n^{\rm B}$,
\begin{align}
\label{eq:e1}\nonumber
{\mathcal{E}}_n(\eta) ={\mathcal{E}}_{n+}+f_n(\eta){\mathcal{E}}_{n-},\;\; {\mathcal{E}}_{n+} = \frac{{\mathcal{E}}_n^{\rm F}+{\mathcal{E}}_n^{\rm B}}{2}\;\;{\mathcal{E}}_{n-} = \frac{{\mathcal{E}}_n^{\rm F}-{\mathcal{E}}_n^{\rm B}}{2}\;,
\end{align}
where $f_n(\eta)$ is an odd function that controls the relative mixture of the eccentricity vectors for the forward and backward going quark participants: $f_n(\infty)=1$ and $f_n(-\infty)=-1$, and ${\mathcal{E}}_{n+}\approx{\mathcal{E}}_{n}$ is the eccentricity calculated using all participants~\footnote{We find that $\varepsilon_{n+}$ is larger than $\varepsilon_{n}$ by upto 20\% in mid-central collisions in large system due to the difference in center-of-mass locations for forward-going and backward-going nucleons (see Ref.~\cite{Jia:2014ysa}).}. Note that the ${\mathcal{E}}_{n\pm}$ fluctuate event to event but are constants within an event. Assuming $f_n(\eta)$ in each event is a slowly varying function near mid-rapidity, Ref.~\cite{Jia:2017kdq} shows that
\begin{align}
r^{\rm s}_{n}(\eta) \approx 1-2\eta a_n A_n\;, a_n =\lr{\frac{\partial f_n}{\partial \eta}|_{\eta=0}f_n(\Eref)}\;,\; A_{n} \equiv \frac{\lr{\varepsilon_{n-}^{2}}}{\lr{\varepsilon_{n+}^{2}}+\lr{\varepsilon_{n-}^{2}}}\approx \frac{\lr{\varepsilon_{n-}^{2}}}{\lr{\varepsilon_{n}^{2}}}\;.
\end{align} 
where $a_n$ is a constant that encodes information about the $f_n(\Eref)$, and $A_n$ controls the strength of the eccentricity decorrelations. 

In the linear response picture, the flow harmonics are driven by the overall eccentricity:
 \begin{align}\label{eq:4a}
\sqrt{\lr{v^2_n}} &=\kappa_n \sqrt{\lr{\varepsilon^2_n}}\;,
\end{align}
where we use the fact that harmonic flow can only be measured via the two-particle correlation method which corresponds to $\lr{v^2_n}$. The response coefficient $\kappa_n$ captures the effects of the viscous damping and depends mainly on the overall size of the system ($\npart$ or number of quark participant $\nqp$). With similar argument, we hypothesize that flow decorrelations should be driven by eccentricity decorrelations,
 \begin{align}\label{eq:4b}
F_n &=\kappa_n'A_{n}\;.
\end{align}
The coefficients $\kappa_n' \approx a_n$ is controlled by the mixing function $f_n(\eta)$, whose dependences on centrality is currently unknown. Furthermore, although the influence of $\kappa_n$ to $A_n$ is expected to largely cancel between the numerator and denominator, some residual dependence could remain since the $\kappa_n$ for $\varepsilon_{n}^{\mathrm{F}}$ and $\varepsilon_{n}^{\mathrm{B}}$ can be different if $\npartf\neq\npartb$ in a given event.

In studying the system-size dependence, it is useful to consider ratios of flow harmonics or flow decorrelations as a function of $\npart$ or $\npart/2A$ where $A$ is the atomic number,
\begin{align}\label{eq:10}
\frac{v_n^{\mathrm{A+A}}}{v_n^{\mathrm{B+B}}}=\frac{\kappa_n^{\mathrm{A+A}}}{\kappa_n^{\mathrm{B+B}}} \frac{\varepsilon_n^{\mathrm{A+A}}}{\varepsilon_n^{\mathrm{B+B}}} \,\, ,
\frac{F_n^{\mathrm{A+A}}}{F_n^{\mathrm{B+B}}}=\frac{\kappa_n^{'\mathrm{A+A}}}{\kappa_n^{'\mathrm{B+B}}} \frac{A_n^{\mathrm{A+A}}}{A_n^{\mathrm{B+B}}} \,\, , \,\,
\end{align}
The $\npart$ is a proxy for absolute system size while $\npart/2A$ can be considered as a measure for scaled system size. When plotted as a function of $\npart$, the $\kappa_n$ is expected to cancel in the $v_n$-ratio and $v_n^{\mathrm{A+A}}/v_n^{\mathrm{B+B}}\approx \varepsilon_n^{\mathrm{A+A}}/\varepsilon_n^{\mathrm{B+B}}$. In contrast, for the same $\npart/2A$, the longitudinal structure of the initial state is expected to have similar F-B asymmetry in the number of sources and similar $f_n(\eta)$~\footnote{In the limit of many sources per-nucleon or optical Glauber, the F-B asymmetry should be a universal function of $\npart/2A$.}, and therefore the $F_n$-ratio is expected to approximately scales with $A_n$-ratio, i.e. $F_n^{\mathrm{A+A}}/F_n^{\mathrm{B+B}}\approx A_n^{\mathrm{A+A}}/A_n^{\mathrm{B+B}}$. The Eq.~\ref{eq:10} and above arguments are the main assumptions used in this paper for our predictions of the system-size dependence of the eccentricity decorrelations.

The eccentricity and its decorrelations are calculated using a standard quark Glauber model from Ref.~\cite{Loizides:2016djv}. Three quark constituents are generated for each nucleon according to the ``mod'' configuration~\cite{Mitchell:2016jio}, which ensures that the radial distribution of the three constituents after recentering follows the proton form factor $\rho_{\mathrm{proton}}(r) = e^{-r/r_0}$ with $r_0=0.234$~fm~\cite{DeForest:1966ycn}. The nucleons are assumed to have a hard-core of 0.4 fm in radii, their density distribution is given by the Woods-Saxon profile,
\begin{align}\label{eq:1}
\rho(r)=\frac{\rho_0}{1+e^{(r-R_0)/a}}\;,
\end{align}
where, $\rho_0$ is the nucleon density, $R_0$ is the nuclear radius and $a=0.55$ is the skin depth. The value of quark-quark cross-section is chosen to be $\sigma_{\mathrm{qq}}=18$~mb, which corresponds to nucleon-nucleon inelastic cross-section $\sigma_{\mathrm{nn}}=68$~mb for $\sqrtsnn=5.02$ TeV. For the study of system size dependence, six spherical nuclei are considered, see Table \ref{tab:1}. The effect of the deformation are considered for Xenon and Uranium, denoted by $\textrm{Xe}^d$ and $\textrm{U}^d$, according to,
\begin{align}\label{eq:3}
\rho(r,\theta)=\frac{\rho_0}{1+e^{(r-R_0(1+\beta_2Y_{20}(\theta)+\beta_4Y_{40}(\theta)))/a}}
\end{align}
where $Y_{20}$ and $Y_{40}$ are Legendre polynomials and $\beta_2$ and $\beta_4$ are deformation parameters.  The deformation parameters are chosen as $\beta_2=0.28$ and $\beta_4=0.093$ for $\textrm{U}^d$~\cite{Masui:2009qk} and $\beta_2=0.162$ and $\beta_4=-0.003$ for $\textrm{Xe}^d$~\cite{Moller:2015fba,Giacalone:2017dud}.
\begin{table}[h!]
\centering
\begin{tabular}{|c|c|c|c|c|c|c|}
\hline
System        & U    & Pb   & Au   & Xe  & Zr   & Cu  \\\hline
Atomic number & 238  & 208  & 197  & 129 & 96   & 63  \\\hline
$R_0$ (fm)    & 6.81 & 6.62 & 6.38 & 5.42& 5.08 & 4.20\\\hline
\hline
\end{tabular}
\caption{A list of nuclei used in this study.}
\label{tab:1}
\end{table}

The Glauber simulation is performed for various collision systems to generate the positions of the participant nucleons and quark constituents, which are used to calculate eccentricity $\varepsilon_n$ and eccentricity decorrelations $A_n$. The eccentricity vector is calculated using the transverse positions of quarks as $\mathcal{E}_n = -\lr{r^n e^{\mathrm{i}n\phi}}/\lr{r^n}$. Similarly, the $\mathcal{E}_{n}^{\mathrm{F}}$ and $\mathcal{E}_{n}^{\mathrm{B}}$ are calculated using only the forward-going and backward-going quarks, respectively, which are then used to obtain the $A_n$.

\vspace*{-0.3cm}\section{Result}\label{sec:3}
\begin{figure}[h!]
\begin{center}
\includegraphics[width=0.6\linewidth]{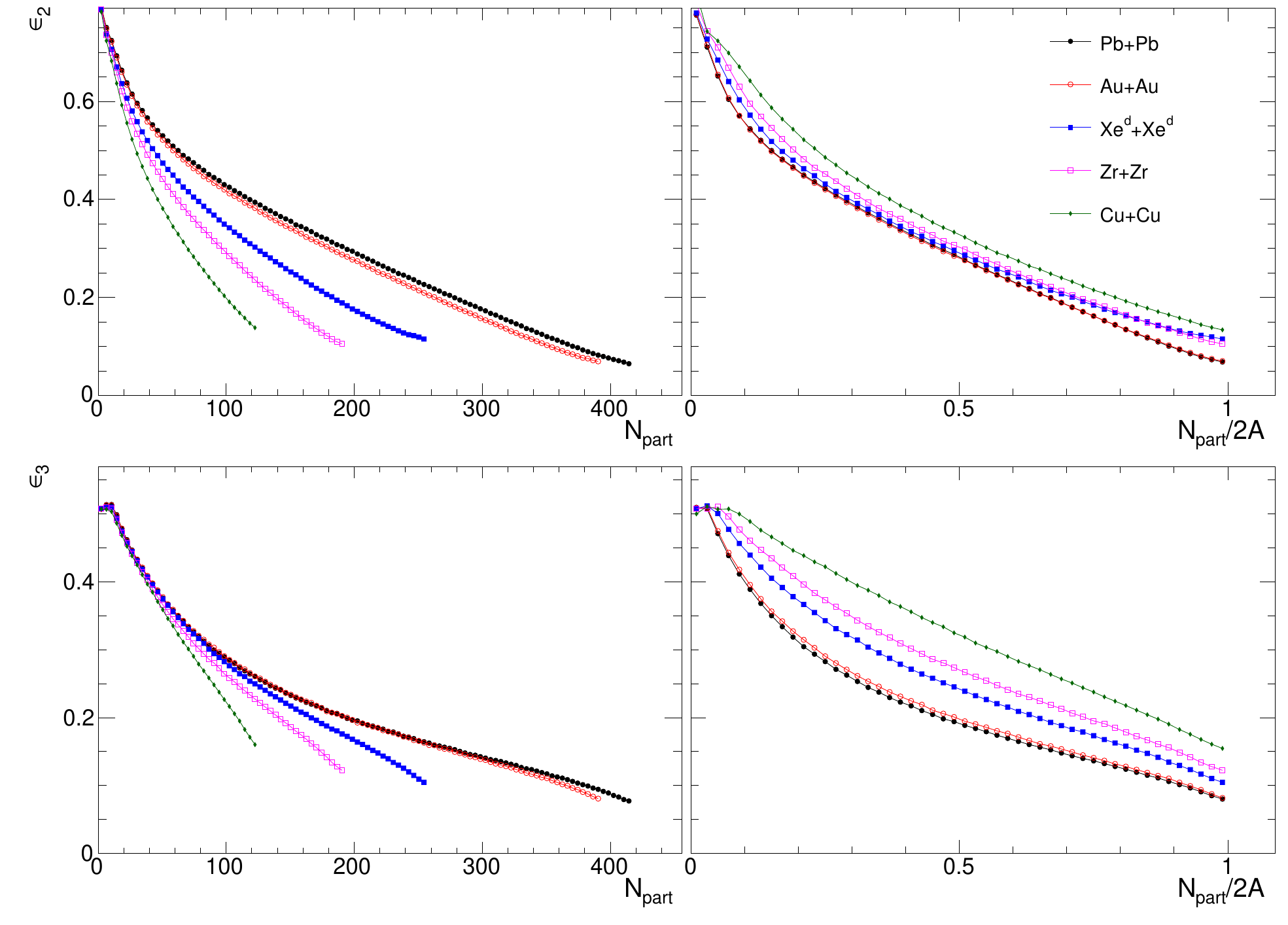}
\end{center}
\vspace*{-0.6cm}
\caption{\label{fig:1} The $\npart$ (left) or $\npart/2A$ (right)  dependence of $\varepsilon_2$ (top) and $\varepsilon_3$ (bottom) for five different symmetric collision systems.}
\vspace*{-0.5cm}
\end{figure}

The top panels of Fig.~\ref{fig:1} show the $\varepsilon_2$ calculated in different collision systems. A clear hierarchy is observed when $\varepsilon_2$ is plotted as a function of $\npart$. However, when plotted as a function of $\npart/2A$, the $\varepsilon_2$ values for different systems nearly collapse on a common curve that simply reflects the centrality-dependent shape of the elliptic geometry of the overlap region. The bottom panels of Fig.~\ref{fig:1} show the results for $\varepsilon_3$. The $\varepsilon_3$ values from different systems overlap at small $\npart$ region, but deviate from each other at larger $\npart$ values. This behavior suggests that although the $\varepsilon_3$ is driven by the random fluctuations of quark constituents, it results in common $\varepsilon_3$ values only when $\npart$ is not too large. In the large $\npart$ region, the $\varepsilon_3$ also depends on the size and the $\varepsilon_2$ of the overlap region (for example due to the anti-correlation between $\varepsilon_2$ and $\varepsilon_3$~\cite{Huo:2013qma}). 

\begin{figure}[h!]
\begin{center}
\includegraphics[width=0.6\linewidth]{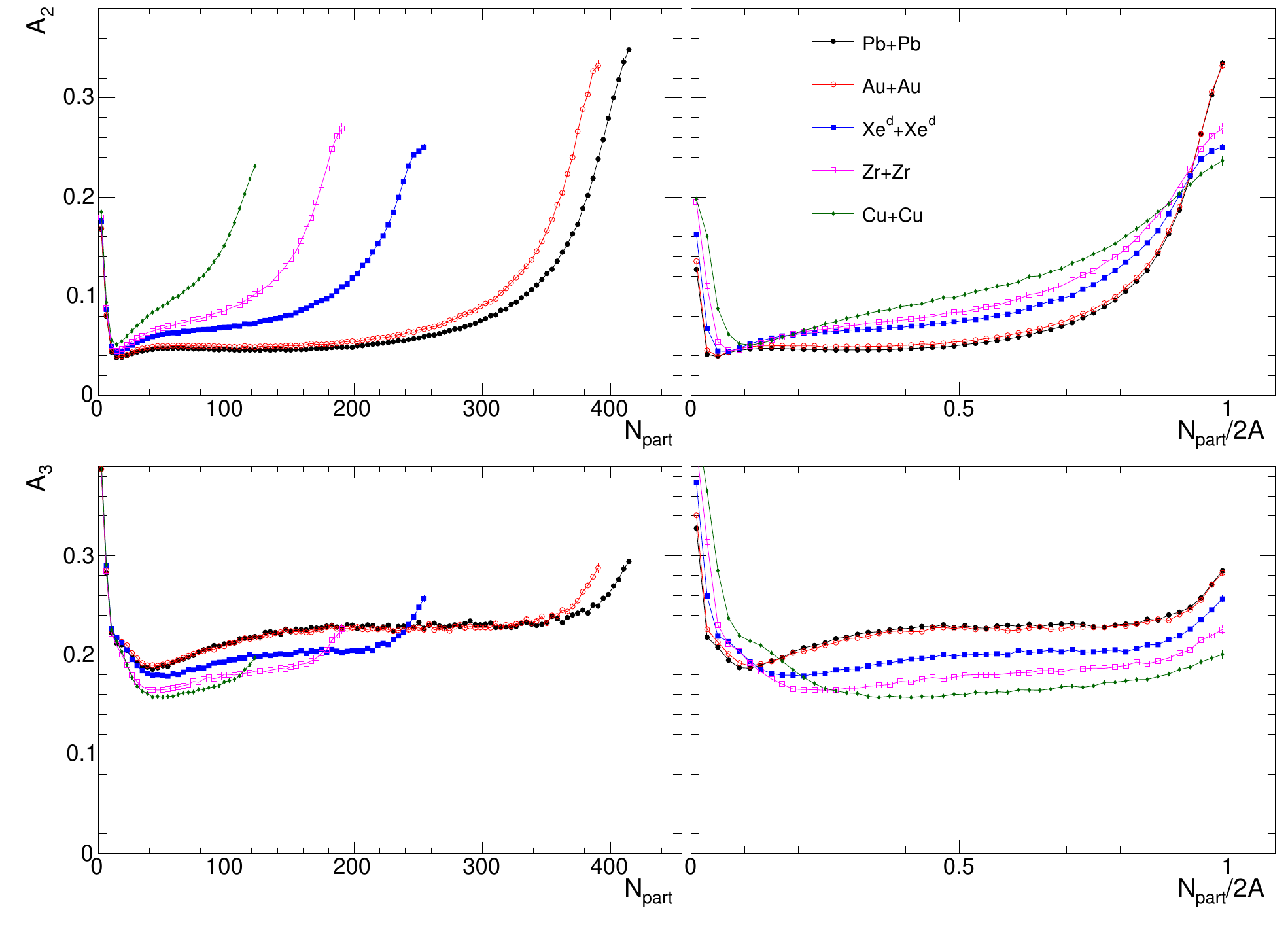}
\end{center}
\vspace*{-0.6cm}
\caption{\label{fig:2}  The $\npart$ (left) or $\npart/2A$ (right)  dependence of eccentricity decorrelations $A_2$ (top) and $A_3$ (bottom) for five different symmetric collision systems.}
\vspace*{-0.5cm}
\end{figure}

The left panels of Fig.~\ref{fig:2} show the results of eccentricity decorrelations $A_2$ and $A_3$ as a function of $\npart$. The $A_2$ values are larger for small systems, while the opposite trend is observed for the $A_3$. This opposite system-size dependence trend between $A_2$ and $A_3$ is much more obvious when they are plotted as a function of $\npart/2A$ in the right panels. 

\begin{figure}[h!]
\begin{center}
\includegraphics[width=0.6\linewidth]{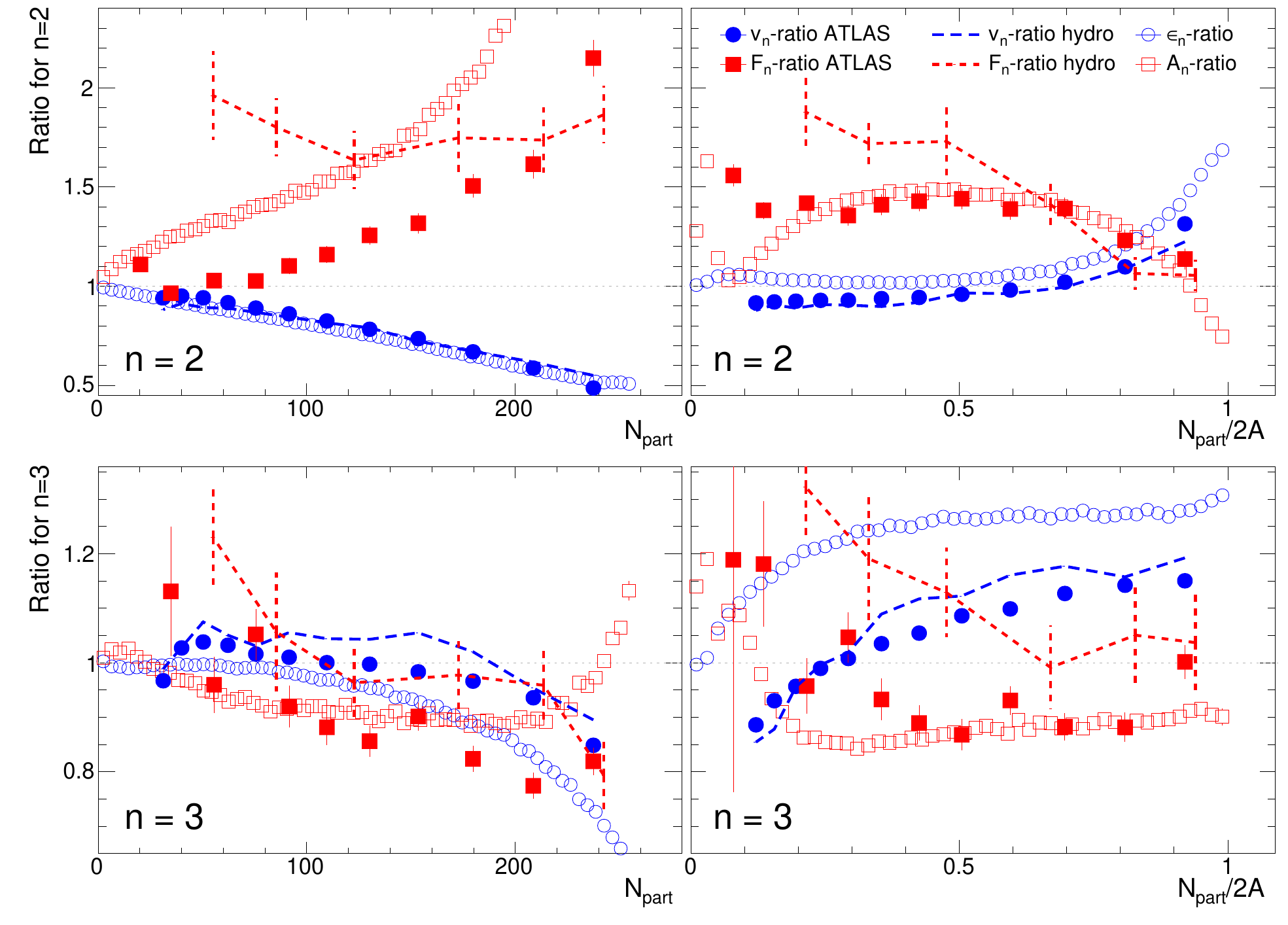}
\end{center}
\vspace*{-0.8cm}
\caption{\label{fig:3} The $\npart$ (left) or $\npart/2A$ (right) dependence of $\varepsilon_n$-ratio and $A_n$-ratio between Xe+Xe and Pb+Pb collisions for $n=2$ (top) and $n=3$ (bottom). They are shown for data (solid symbols), Glauber model (open symbols). They are compared also with hydrodynamic model predictions (lines) for $v_n$-ratio~\cite{Giacalone:2017dud} and $F_n$-ratio~\cite{Pang:2018zzo,Wu:2018cpc}.}
\end{figure}

Recently, the ATLAS Collaboration has performed the first measurement of the system-size dependence of flow decorrelations for $v_2$ and $v_3$~\cite{Aad:2020gfz}. We can check how well the Glauber model describes the change between Xe+Xe and Pb+Pb observed in the ATLAS data. The left panels of Fig.~\ref{fig:3} show the $\npart$ dependence of $\varepsilon_n$-ratios and $A_n$-ratios, and they are compared with the $v_n$-ratios and $F_n$-ratios respectively from the data and hydrodynamic model predictions. The $\varepsilon_n$-ratios agree with the $v_n$-ratio very well and this is because the response coefficient $\kappa_n$ depend only on the overall size of the overlap region described by $\npart$, and therefore cancel in the ratios. On the other hand, the $A_n$-ratios show qualitatively similar trends as the $F_n$-ratios, but are quantitatively different especially for $n=2$. Note that the hydrodynamic model predictions reproduce the $v_n$-ratios but fail to describe the $F_n$-ratios, implying the model does not have the correct initial-state condition in the longitudinal direction. 

The right panel of Fig.~\ref{fig:3} show the same ratios calculated as a function of $\npart/2A$. It is clear that $\varepsilon_n$-ratios do not describe the $v_n$-ratios due to the fact that the $\kappa_n$ do not cancel, which lead to about 10\% difference between $\varepsilon_n$-ratio and $v_n$-ratio for $n=2$ and 10--20\% difference for $n=3$. However, the $A_n$-ratios, which are expected to be relatively insensitive to $\kappa_n$, show an overall good agreement with the $F_n$-ratios. This agreement implies that the $f_n(\eta)$ function controlling the mixing between forward-going and backward-going sources is mostly a function of centrality percentile or $\npart/2A$ between different systems. This result also supports the opposite hierarchy between the system-size dependence of $A_2$ and the system-size dependence of $A_3$ in Fig.~\ref{fig:2}.

The deformation of colliding nuclei is known to influence the $\npart$ dependence of $\varepsilon_n$ and $v_n$~\cite{Filip:2009zz,Rybczynski:2012av,Adamczyk:2015obl}.
\begin{figure}[h!]
\begin{center}
\includegraphics[width=0.6\linewidth]{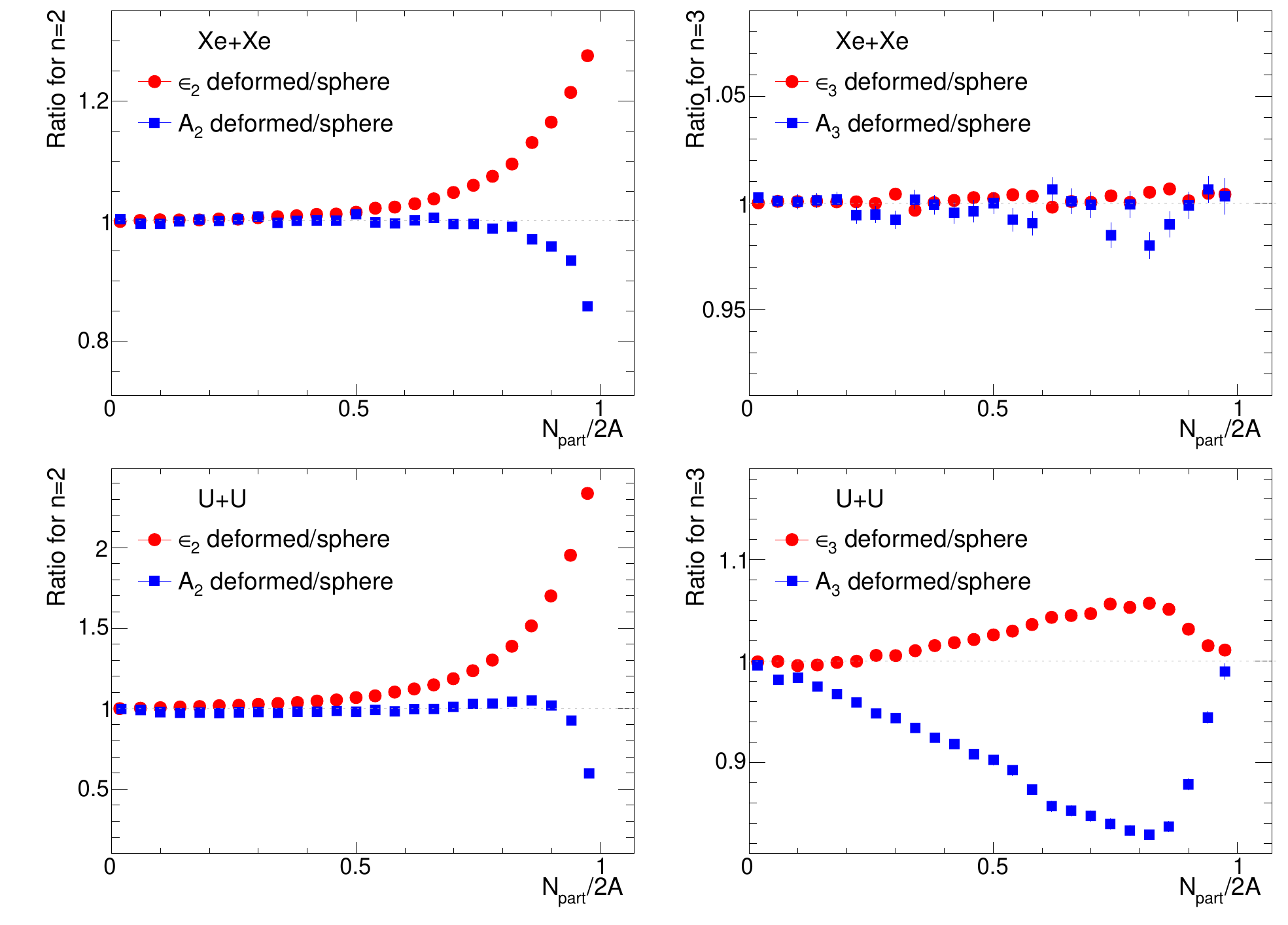}
\end{center}
\vspace*{-0.8cm}
\caption{\label{fig:4} The $\npart$ dependence of $\varepsilon_n$-ratio and $A_n$-ratio for $n=2$ (left panels) and $n=3$ (right panels) for Xe+Xe (top) or U+U (bottom) between with and without nuclear deformation from Glauber model.}
\end{figure}
An interesting question is whether the eccentricity decorrelations are also affected. Figure~\ref{fig:4} shows the $\varepsilon_n$-ratios and $A_n$-ratios for Xe (top panels) and U (bottom panels) with and without deformation. 
 In the case of Xe, the deformation influences the $\varepsilon_2$ and $A_2$ in central collisions but in the opposite direction, i.e. deformation increases the $\varepsilon_2$ but reduces the $A_2$. The deformation has very little influences on the $\varepsilon_3$ and $A_3$. In the case of U, the deformation increases $\varepsilon_2$ over a broader centrality range. The influence on $A_2$ is a bit non-trivial: the deformation has little effect in the mid-central and peripheral collisions, but decreases the $A_2$ in the ultra-central collisions. The deformation increases the values of $\varepsilon_3$ but decreases the values of $A_3$. These features can be searched for in the experimental analyses, for example by comparing the Ru+Ru and Zr+Zr at RHIC which have the same atomic number but different amount of deformation~\cite{Deng:2016knn}.

\begin{figure}[h!]
\begin{center}
\includegraphics[width=0.6\linewidth]{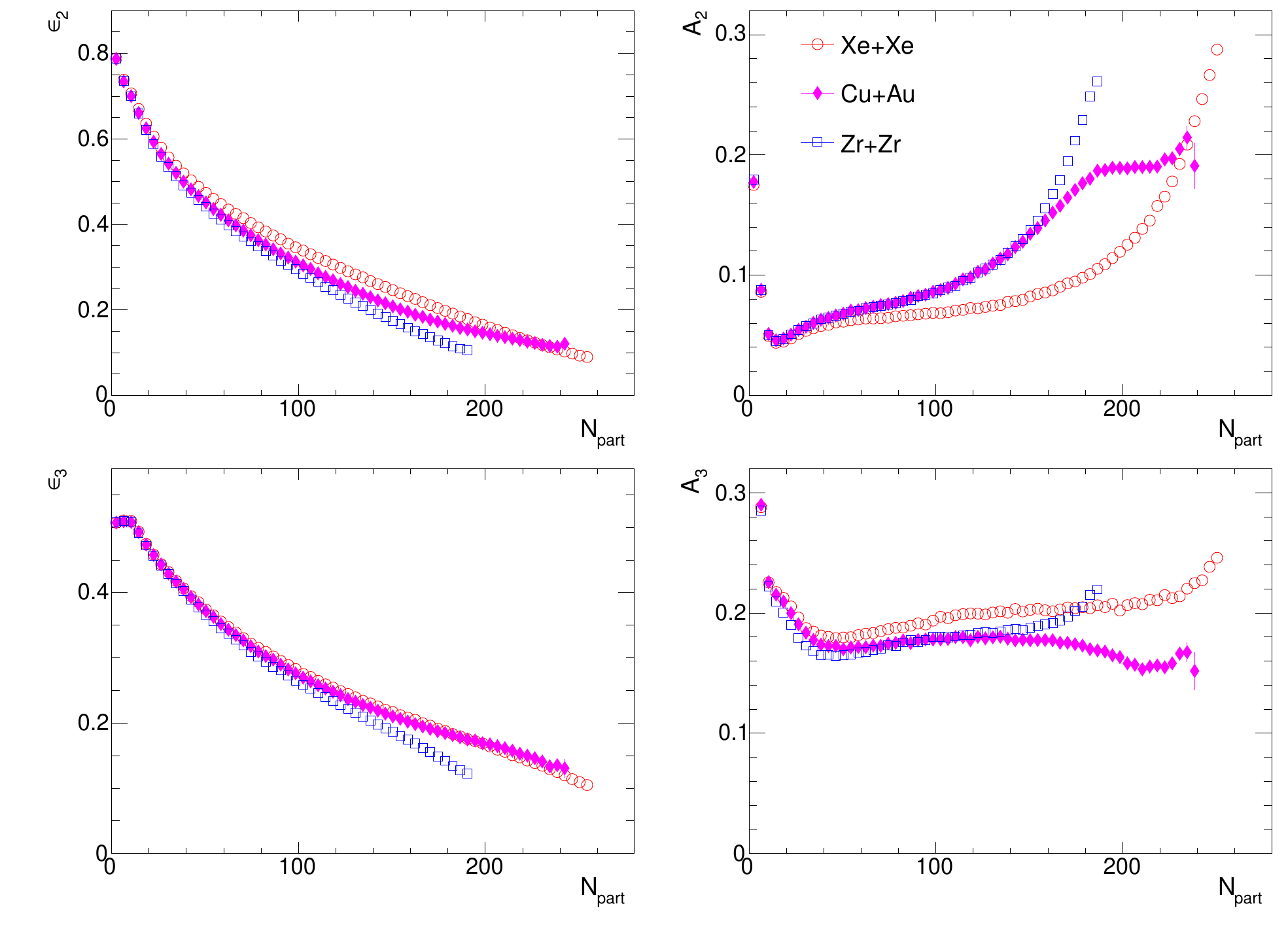}
\end{center}
\vspace*{-0.5cm}
\caption{\label{fig:5} The $\npart$ dependence of $\varepsilon_2$ (top-left), $\varepsilon_3$ (bottom-left), $A_2$ (top-right) and $A_3$ (bottom-right) compared between asymmetric Cu+Au collision system and symmetric Zr+Zr and Xe+Xe systems.}
\end{figure}

Figure~\ref{fig:5} shows our prediction of the eccentricity decorrelations in asymmetric collision system Cu+Au, for which the intrinsic asymmetry between the $\npartf$ and $\npartb$ should also influence the behavior of $\varepsilon_n$ and $A_n$. Since the overall system size for Cu+Au is in between Zr+Zr and Xe+Xe, we compare the $\varepsilon_n$ and $A_n$ among these three systems. The $\npart$ dependences of $A_n$ in central Cu+Au region are distinctly different from those in the Zr+Zr and Xe+Xe systems. This is because over a wide range in the central collisions region, all the nucleons from Cu participate in the collisions, while the nucleon participants in Au still increases, resulting in a weak dependence of both $\varepsilon_n$ and $A_n$ on the $\npart$. 

\vspace*{-0.3cm}\section{Summary}
We discussed the dependence of elliptic flow $v_2$ and triangular flow $v_3$ and their longitudinal decorrelation coefficients $F_2$ and $F_3$ on the choice of collision systems in terms of the size, deformation and asymmetry of the nuclei with atomic number $A$. Hydrodynamic model simulation shows that the harmonic flow are driven by the initial-state eccentricity, $\varepsilon_n$ $v_n\propto \varepsilon_n$, and the flow decorrelations are directly determined by the eccentricity decorrelations in the longitudinal direction $A_n$, $F_n\approx A_n$. We estimate the values of $\varepsilon_n$ and $A_n$ in various collision systems using a Monte-Carlo quark Glauber model, which assumes three constituent quarks for each nucleon in determining the initial state, and the results are presented as a function of number of nucleon participants $\npart$ or that normalized by the total number of nucleons of the collision systems $\npart/2A$. 

We found that the $A_2$ is larger for smaller collision systems, while the opposite ordering is observed for the $A_3$. The ratios of $\varepsilon_n$ or $A_n$ between Xe+Xe and Pb+Pb are compared with the ratios of $v_n$ or $F_n$ measured by the ATLAS Collaboration as a function of both $\npart$ and $\npart/2A$. The $\varepsilon_n$-ratios approximately agree with $v_n$-ratios as a function of $\npart$, while the $A_n$-ratios agree with $F_n$-ratios as a function of $\npart/2A$. This behavior is consistent with our understanding that the flow response coefficient $\kappa_n=v_n/\varepsilon_n$ depend on the overall system size described by the $\npart$, while the coefficient for flow decorrelations $F_n/A_n$ might depend only on the overall shape of the overlap region controlled by the $\npart/2A$. Current hydrodynamic models fail to describe simultanously the flow decorrelations in Xe+Xe to Pb+Pb, and this failure implies that their initial longitudinal structure based on lund-string picture of AMPT nees improvement.

We further compared the $\varepsilon_n$ and $A_n$ for Xe and U nuclei with and without the effects of nuclear deformation. For the modest deformation parameter $\beta_2=0.162$ of Xe, the deformation affects the $\varepsilon_2$ and $A_2$ only in ultra-central collisions. For the large deformation parameter $\beta_2=0.28$ of U, the nuclear deformation influences both the $n=2$ and $n=3$ of $\varepsilon_n$ and $A_n$ over a broad centrality range. The deformation always increases the value of $\varepsilon_n$ while decreases the value of $A_n$. These features might be searchable using the Zr+Zr and Ru+Ru isobar data from the STAR Collaboration since Zr and Ru are expected to have slightly different $\beta_2$ values~\cite{Deng:2016knn}. We also considered the Cu+Au asymmetric collision system, which shows a $\npart$ dependence of $\varepsilon_n$ and $A_n$ different from symmetric systems in the central region.

This paper serves as an exploratory study of the possible influence of various nuclear geometry effects on the harmonic flow and longitudinal flow decorrelation in heavy ion collisions. More quantitative predictions would require coupling the 3D initial condition with state-of-art hydrodynamic model simulation. The final-state effects are expected to significantly change the relation between the $v_n$ and $\varepsilon_n$ in a system-dependent manner, but they may not change too much the relationship between the $F_n$ and $A_n$ as predicted by our calculations.

This research is supported by National Science Foundation under grant number PHY-1613294 and PHY-1913138 (JJ and AB) and by China Postdoctoral Science Foundation 2019M662319 (MN).

\bibliography{decor}{}
\bibliographystyle{apsrev4-1}

\end{document}